\documentstyle[editedvolume]{crckapb}

\begin{opening}
\title{THE DISTANCES OF THE MAGELLANIC CLOUDS}

\author{ALISTAIR R. WALKER}
\institute{Cerro Tololo Inter-American Observatory, NOAO,\\
  Casilla 603, La Serena, Chile}

\end{opening}

\runningtitle{THE MAGELLANIC CLOUDS}

\begin{document}

\begin{abstract}

The present status of our knowledge of the distances to the
Magellanic Clouds is evaluated from a post-Hipparcos perspective.
After a brief summary of the effects of structure, reddening, age,
and metallicity, the primary distance indicators for the Large 
Magellanic Cloud are reviewed:  The SN 1987A ring, Cepheids, RR
Lyraes, Mira variables, and Eclipsing Binaries.  Distances 
derived via these methods are weighted and combined to produce
final {\it best} estimates for the Magellanic Clouds distance moduli.

\end{abstract}

\section{Introduction}

The distances of the Magellanic Clouds (MC), in particular that for the Large Magellanic
Cloud (LMC), are of great importance for three principal reasons:  Firstly,
the MC are sufficiently populous that they contain many different types of
distance indicator, and are close enough so that most can be measured with high accuracy,
thus they are invaluable for consistency comparisons.
Secondly, the MC are sufficiently remote so that to 
first order the constituents of each may be taken to be at constant distance 
from us.  Finally, the MC are a sanity check, for if we cannot agree
upon the distance to two galaxies that are only a few 10's of kpc distant
from us, how can we be sure of distances to more remote galaxies?

For these reasons the literature discussing MC distances
is large.  Chapter 2 of Westerlund (1997) comprehensively summarizes the subject to 
that date,  somewhat disconcertingly he finishes with the
statement {\em In view of all the problems involved in the distance determinations
it is necessary to admit that the distances of the two Clouds are still not
sufficiently well known}.  In this chapter we will investigate whether or not the
situation has changed for the better in the intervening two years, during
which time Hipparcos astrometry has become available and photometry of massive
numbers of stars as a byproduct of microlensing surveys is appearing in the
literature.

We will begin (2) by reviewing some properties of the MC relevant to the present 
investigation, follow (3) by considering various distance indicators
and conclude (4) by summarizing the present status of the MC distances together
with some indication of possible future improvements.  Almost all of the  
standard candles useful for MC distances rely on a galactic calibration, which 
for many of them will be discussed elsewhere in this volume.  

\section{Relevant Properties of the Magellanic Clouds}

\subsection {structure}

The LMC is a barred spiral, and defines the {\em Magellanic} subclass (SBm), 
(de Vaucouleurs \& Freeman 1972, 
Wilcots et al. 1996).  Kinematical studies (Olszewski et al. 1991, Schommer 
et al. 1992) 
show that even the oldest populations (eg globular
clusters containing RR Lyraes) have disk kinematics and there is presently
no evidence for a pressure-supported halo (Olszewski, Suntzeff \& Mateo 1996).
Thus for the LMC a satisfactory
assumption is that most constituents are confined close to a well-defined disk 
that is only mildly tilted with respect to the plane of the sky. 

However both the LMC and especially the Small Magellanic Cloud (SMC) 
show evidence for interactions between 
each other and the Galaxy.  The extension of the SMC is very considerable
 in the line-of-sight (Gardiner \& Hatzidimitriou 1992)  
thus, unless the particular component of the
SMC corresponding to a potential distance indicator can be unambiguously identified,
it is of little use for finding the mean distance to the SMC.  Consequently
most of what follows pertains to the LMC rather than the SMC, and perhaps the major utility 
of the SMC is as an aid in testing the metallicity sensitivity of distance indicators.

\subsection {reddening}

The main bodies of the MC are between galactic latitudes $-29^{\circ}$ to $-35^{\circ}$ (LMC)
and $-43^{\circ}$ to $-46^{\circ}$ (SMC), and foreground reddening is expected to be 
non-negligible, especially for the LMC.   Tanvir (1996) emphasizes that for
certain distance scale applications having specific knowledge of the reddening can be
circumvented, while in addition reddening corrections in the infrared are much
less significant than in visual passbands.    Bessell (1991) re-evaluated the
foreground and internal reddening for both the SMC and LMC.  He concludes that
the foreground reddening for the SMC shows little variation and probably lies
between $E(B-V) = 0.04$ and 0.06, while the foreground to the LMC is more varied,
$E(B-V) = 0.04$ to 0.09.  Average reddening within the SMC is about $E(B-V) = 0.06$,
with a similar figure for the LMC but he finds for the latter that there is a wider
range of values than for the SMC.  More recent
observations for the LMC confirm this picture.
Schlegel, Finkbeiner \& Davis (1998) present a full-sky galactic 
extinction map that is twice as accurate as the much-used Burstein \& Heiles
(1982) map in regions of low to moderate reddening.  The MC (and M31) are 
not removed from the new map but unfortunately accurate reddenings through these 
galaxies is
not possible.  The typical foreground reddening, measured from dust emission in
surrounding annuli, is $E(B-V) = 0.075$ for the LMC and $E(B-V) = 0.037$ for the
SMC.  Noteworthy is that the new map is offset, with 0.02 mag higher reddening in
high galactic latitudes, compared to Burstein \& Heiles (1982).

A major study of the reddening foreground to the LMC is that 
of Oestreicher, Gochermann \& Schmidt-Kaler (1995), who
from UBV colors of 1409 galactic stars derive a reddening map with resolution 10
arcmin.  The mean reddening is $E(B-V) = 0.06 \pm 0.02$, 
however the distribution appears quite clumpy with a range from
$E(B-V) = 0.0$ to 0.15.  The higher values are interpreted as corresponding 
to dust clouds in the solar vicinity, these project to diameters typically 30-60 
arcmin.  

The reddening internal to the LMC is treated by Oestreicher \& Schmidt-Kaler (1996) 
from UBV photometry and spectral classifications.  Their map of the reddening 
distribution correlates quite well with the HI column density (Luks \& Rohlfs 1992)
and the IRAS 25 micron emission map (Israel \& Schwering 1986).  The highest
reddening occurs in the regions of 30 Doradus and the supershell LMC 2, reaching a
maximum of $E(B-V) = 0.29$, and stars in the bar are in general more highly reddened
than elsewhere.  However there are highly reddened stars spread out over the LMC, 
and conversely stars in the bar with rather low reddening.  Some of the former stars, 
all of very high luminosity, may be reddened by circumstellar material, while some 
proportion of the latter 
stars may be located on the near side of the bar and thus suffer little internal
LMC reddening.   They also find that there are strong selection effects
in that their intrinsically fainter stars ($V_0 > 13.3$) show low reddenings, with
median approximately $E(B-V) = 0.07$.

The general conclusions are that:\\
1) Magnitude-limited samples are going be the lower-reddened stars.\\
2) The reddening is patchy enough, both galactic and in the LMC, that 
proceeding on a star by star or cluster by cluster basis seems prudent,
if it is possible.   In this context, 
an important program is that of Madore, Freedman \& Pevunova (in preparation) where
OB star reddenings are being determined in the line of sight to many of the
LMC Cepheid calibrators.  This will allow direct comparison with the semi-empirical
reddenings determined by Caldwell \& Coulson (1986) for most of these stars.\\
3) The more outer parts of the LMC suffer only galactic foreground reddening, as
evinced by reddenings found for several LMC clusters containing RR Lyraes (Walker 
1992). \\
4) The median reddening for stars in the LMC is $E(B-V) \sim 0.10$ and for those
in the SMC probably slightly less, $E(B-V) \sim 0.08$.\\
5) With $A_K \sim 0.03$ on average, the advantages of working in the infrared are 
obvious.

\subsection {age, metallicity}

Most of the distance estimates for the MC rely on comparing  a sample of stars
in the MC with a corresponding galactic sample for which we know individual distances. 
In general, as might be expected, the galactic samples are rather better defined 
in terms of age and metallicity than are the MC samples.  In some cases, for example 
Cepheids in galactic open clusters, the calibrating sample is rather sparse.
Stated in general terms, we first need to have a reliable calibration for the
galactic sample of stars,  then we need to make the relevant observations for
the MC sample, then we have to determine any differences between the samples
and apply a correction if this affects the distance.  Obviously, any direct
(eg geometric) distance measurement circumvents these problems.  
Two illustrative examples follow.  

Firstly, the sizeof the metallicity dependency of the Cepheid Period-Luminosity 
(PL) relation has been controversial for several years
(Madore \& Freedman 1991, Gould 1994a).  Observational data relevant to the
question include metallicities for galactic Cepheids in open clusters (Fry 
\& Carney 1997), for MC Cepheids (Luck et al. 1998), discussions of Cepheid data 
from microlensing project databases (Beaulieu et al. 1995), and the 
Hubble Space Telescope (HST) 
extra-galactic Cepheid surveys Freedman et al. 1994, Saha et al. 1994).
These new results have been analyzed (Sasselov et al. 1997, Kochanek 1997, Kennicutt
et al. 1998) with particular attention devoted to the V and I bands used for both
HST surveys,  where typically   (Kennicutt et al. 1998) an effect on Cepheid
distance moduli from $V$ and $I$ PL relations is $\sim -0.25 \pm 0.25$ mag/dex, although
at times a stronger dependence has been suggested (Gould 1994a, Sekiguchi \&
Fukugita 1998).  With
mean $[Fe/H] = -0.34 (\sigma = 0.15, n=32)$ for the LMC, and $[Fe/H] = -0.68 
(\sigma = 0.13, n=25)$ for the SMC (Luck et al. 1998), the metallicity corrections
are not large, particularly for the LMC.  These Cepheid metallicities are in the
mean more metal poor than earlier empirical or semi-empirical estimates (eg Caldwell
\& Coulson 1986, Laney \& Stobie 1994), where $[Fe/H] = -0.15$ was adopted for the 
LMC. Note that a change from $[Fe/H] = -0.15$ to $-0.3$ will change the mean reddening
for the LMC found by Caldwell \& Coulson (1986) from $E(B-V) = 0.074$ to 0.059, with
consequent affect on use of this data for determining distances (Feast 1998). 

Secondly, a new distance indicator, I magnitudes of red clump stars, has been developed 
by Paczy\'{n}ski and Stanek (1998).  This can be directly calibrated from Hipparcos
parallaxes, from which the mean absolute magnitude of 228 solar neighborhood red 
clump stars is found to be $M_{I} = -0.23 \pm 0.03$.  
Local group galaxies distances can thus be directly calibrated from Hipparcos
results in a single step.  Udalski et al. (1998) find MC distance moduli of 18.56
and 18.08 for the SMC and LMC respectively, with very small statistical errors (0.03 
mag).  Despite the seemingly photometrically well-defined populations of 
clumps stars locally and in the MC, these MC distance moduli are disconcertingly
short compared to all other indicators.  Cole (1998) suggests that a luminosity dependence
of the red clump stars on both age and metallicity may resolve the apparent
discrepancy, and revises the distances to the LMC and SMC to $18.36 \pm 0.17$
and $18.82 \pm 0.20$ respectively.  Similar results are found by Giraldi et al.
(1998).  This method has much potential, once 
age and metallicity dependence effects have been definitively settled, given the
ubiquitous nature of red clump stars and consequent small statistical errors in
ensemble mean magnitudes.

\section{MC Distance Calibrations}

\subsection{A Direct Distance to the LMC - The SN 1987A Ring}

Panagia et al. (1991) were the first to determine a direct distance to supernova
 (SN) 1987A by 
comparing high-accuracy measurements of the angular and physical size of the 
circumstellar ring surrounding the SN.  The method is conceptually simple; the
physical size can be calculated by measuring the light travel time to the ring, 
derived from  International Ultraviolet Explorer (IUE) 
lightcurves of UV emission lines (eg NIII], NIV], NV, CIII]) 
observed between days 8 and 700 after the explosion, while the angular size is
directly measured from HST images (eg O[III]) of necessity taken at later epochs.  
As might be expected, several assumptions must be made, and differing interpretations 
of the data are possible.  The assumptions can be listed as:\\
	1) The IUE lightcurves and the more recent images both correspond  
to gas that is in the same physical location. \\ 
	2) The structure visible is a ring, and not some more complicated 
geometry.\\
	3) The caustics in the IUE light curves do indeed represent the extreme 
light travel times.\\
	4) The ring is circular and smooth.\\
	5) The delay time between when the UV pulse first hit the ring and the
appearance of the UV line emission is negligibly small.

In general, quoted errors have reflected fitting errors to the observational data
and have not attempted to realistically account for systematic effects arising 
from incorrect assumptions or inadequate models.

Gould (1994b) discussed assumptions 2), 3), and 4) in some detail, showing that the 
first two did indeed appear to be valid, and that ellipticity
had little effect on the distance.  Crotts, Kunkel, \& Heathcote (1995) confirmed the
ring structure mostly from light-echo data.  It should be noted that proof that the ring is
circular will be tested when the SN shock hits it, a process that appears
to be commencing (Sonneborn et al. 1998).  Assumption 5) is generally thought to 
be valid, but if the delay time is accepted to be indeterminant then the derived
distance becomes an upper limit rather than an equality (Gould 1995).  The initial
assumption 1) has been modeled by Lundquist and Sonneborn (1997), again a conservative 
interpretation of their results gives only an upper limit to the distance.  We shall
now discuss some of these results in more detail. 

All analyses must by necessity use some subset of the IUE light curves, which are
relatively noisy due to the IUE entrance aperture including stars 2 and 3, whose
continua dominated the signal.  Hubble Space Telescope (HST) images of the ring were 
first obtained in August 1990 (Jakobsen et al. 1991) with the Faint Object Camera (FOC) 
and have continued to be taken up to the present, in particular with the Space Telescope
Imaging Spectrograph (STIS) which permits spatially resolved spectroscopy and is ideally
suited for study of the SN environment.

The initial analysis by Panagia et al. (1991) used the early HST images in O[III]
to derive the angular size of the ring ($1.66 \pm 0.03$ arcsec, Jakobsen et al. 1991), 
and fitted a simple model to the time
evolution of the NIII], NIV], NV and CIII] lines observed by IUE.  The model fit to
the NIII] lines is reasonable, but is poorer for the other lines.  The latter have
fewer observations and lower S/N than the NIII] data.  The times at which
the lines first appear was found, in the mean, to be $83 \pm 6$ days, and the maximum
$413 \pm 24$ days, these times being simply related to the size and inclination of
the ring.  The inclination of the ring thus derived was found to be in excellent
agreement with the observed elliptical appearance, assuming that the ring is close
to circular.  The estimated distance to SN 1987A was $51.2 \pm 3.1$ kpc.  

Gould (1994b) re-analysed the data, using a slightly different formalism and applied a
correction to the center of the LMC based on the assumption that the SN lies in the
LMC disk.  The distance to the SN  he found was $53.2 \pm 2.6$ kpc.  Gould (1995), in
a more radical re-analysis of the data, fitted the UV light curves with a model based
on those developed by Dwek \& Felton (1992), thought to be more appropriate for a ring
geometry.  He used only the NIII] and NIV] lines.  The angular size of the ring was
taken from Plait et al. (1995), a value 3 \% higher than used previously.  His fits
to the UV light curves yield times 8 \% smaller than found by Panagia et al. (1991),
$75 \pm 3$ days and $390 \pm 2$ days. 
These two changes both act to reduce the derived distance.  If assumption 5)
above is correct then the distance to the SN is $46.7 \pm 0.7$ kpc. 

Sonneborn et al. (1997) perform a re-reduction of the IUE UV data, 
to provide what are likely to be definitive light curves.  They also determined
caustic timings of $84 \pm 4$ and $399 \pm 15$ days. Lundqvist \& Sonneborn (1997) 
re-analyse the ring geometry in detail, using recent [OIII] 
and [NII] HST images.  In their models, together with those of Lundqvist \& Fransson
(1996) they examine assumption 1) above, to conclude that the innermost parts of the 
[NII] emitting zone best represent the gas which emitted the UV lines.  The corresponding
angular radius of the inner edge of the [NII] emitting zone may be as small as $775 \pm 10$ 
mas, and this, together with their UV light curve timings from Sonneborn et al. (1997)
gives an upper limit to the distance of the SN of $54.2 \pm 2.2$ kpc.  

Panagia et al. (1997), who use the new reductions of the
IUE light curves by Sonneborn et al. (1997), and an extensive set of HST images, repeat 
and improve upon their earlier analysis and derive an absolute size of the ring
$ R_{abs} = (6.17 \pm 0.18) 10^{17} $cm  and an angular size $R_{ang} = 808 \pm 17$
mas, to find a distance to the SN of $50.9 \pm 1.8$ kpc.

Gould and Uza (1998), repeat the earlier analysis (Gould 1995) and also adopt the
Sonneborn et al. (1997) re-evaluation of the UV data.   They find shorter times
for the caustic crossings, $80.5 \pm 1.7$ days and $378.3 \pm 4.8$ days but
with less convincing fits than before.  
With an ellipticity estimate for the ring of $0.95 \pm 0.02$ from A. Crotts, they
derive a distance for the SN of $48.8 \pm 1.1$ kpc, again pointing out that if assumption
5) is invalid then this becomes an upper limit.  They also consider that the scenario
of the initial UV and present optical emission lines coming from different zones
is implausible, but note that such an effect could increase the distance modulus
by up to 8 \%,  this conclusion in part based on the early Plait et al. (1995) 
(pre-CoSTAR) HST imaging data.

It is certain that our knowledge of the ring structure will increase 
dramatically as the ring is
illuminated by passage of the SN shock front, via analysis of HST images
and spectroscopy.  More sophisticated models, both of the ring structure
and of the energy distribution at break-out of the EUV radiation, should
allow a consistent interpretation of all the IUE data and give confidence that
we understand the ring structure.  
By contrast, the very careful re-reduction of the IUE
data by Sonneborn et al. (1997) is unlikely to be improved upon, and thus
represents a basic limitation to the timing accuracy of the caustic crossings.
At this time it is clear that the interpretation
of the present observational data sets (Plait et al. 1995,  
 Panagia et al. 1997, Lundqvist \& Sonneborn 1997, Gould \& Uza 1998) have
still not converged.  The scatter of points near the peak of the N III] 
light-curve (see Gould \& Uza 1998 Figure 2) compared to various fitted models
gives little confidence that the true position of the second caustic crossing is 
known to better than a value $390\pm 15$ days.  If the possible systematic 
effects 1) and 5) above are not significant then the Panagia et al. (1997) value for
the SN distance of $50.9 \pm 1.8$ kpc should be close to the true value.  If the 
situation is more complex than they assume then systematic effects 
could move this distance by up to 4 kpc either way.  

A correction to the rotation center (McGee \& Milton 1966, Bessell, Freeman \&
Wood 1986) of the LMC is necessary.  With  rotation center at $\alpha$(1950) =
5h 21m, $\delta (1950) = -69^{\circ}$ 18', PA of line of nodes $171^{\circ}$, and inclination 
$27^{\circ}$, the plane of the LMC at the position of SN 1987A is 700 pc closer to us than
the LMC center.   Xu, Crotts \& Kunkel (1996) from a light-echo analysis show that the 
large complex of young stars and gas, LH 90 and N157C, lies $\sim 500$ pc in front of SN1987A,
and it seems reasonable to suppose that the former lies very close to the plane,
although recent HI absorption studies (Dickey et al. 1996) indicate that the
velocity structure identified with the disk lies at least partially in front of 30 Doradus.
Spyromilio et al. (1995) by contrast, argue from a light-echo analysis of 3 yr of AAT
plates that the N157C bubble lies behind the SN, and given that the echos represent
material in front of the SN then the latter must lie close to the plane.
Panagia et al. (1991) evaluate HI radial velocity structure (McGee \& Milton 1966,
Radhakrishnan et al. 1972) and interstellar absorption components (Blades 1980) 
in the direction of 30 Doradus, and conclude that approximately two-thirds of the main body
of the LMC is in front of SN 1987A.  With the thickness of the LMC disk taken as 600 pc,  this
distance equal to the scale height of the older stars (Freeman, Illingworth \& Oemler 1983)
then SN 1987A is $\sim 100$ pc behind the plane.   Despite these uncertainties in
the location of the SN with respect to the LMC plane, the
correction is not a large one.
We will adopt a position for the SN of $300 \pm 200$ pc behind the plane and 
thus $400 \pm 200$ pc closer to us than the LMC center.   The LMC distance modulus from
this method is therefore $18.55 \pm 0.07$ (random) $\pm 0.16$ (systematic).

\subsection {Cepheid Distances}

The use of Cepheids as extragalactic distance indicators has recently been
comprehensively reviewed by Tanvir (1996).
Cepheid distances to the MC are traditionally found by comparing  PL 
(period luminosity) or PLC (period luminosity color) relation zeropoints 
between the MC and our galaxy (Feast \& Walker 1987, Laney \& Stobie 1994).
The galactic calibration can be via Cepheids in open clusters and associations,
Cepheids with Baade-Wesselink distances (Gieren, Fouqu\'{e}, \& Gomez 1997) or via 
Cepheids with 
Hipparcos parallaxes (Feast \& Catchpole 1997, Madore \& Freedman 1997).

\subsubsection{Period Luminosity relation}

There are advantages to specifying the slope and zeropoint of the Cepheid period 
luminosity in the infrared,
where the narrower width of the instability strip produces a tighter
PL relation, pulsation amplitudes are smaller, and effects of reddening less, when compared to 
visual bandpasses.  The galactic calibration has traditionally proceeded from
a zero age main sequence (ZAMS) 
calibration of the few galactic clusters and associations containing Cepheids,
with slope defined from the LMC Cepheids and zeropoint ultimately anchored to the
distance to the Hyades.  In recent times the zeropoint calibration has also been referenced
to the Pleiades, with distance derived from a fit of the Pleiades ZAMS to local parallax
stars (van Leuuwen 1983), on the grounds that the Pleiades is more similar in age and metallicity
to the Cepheid-containing clusters.  With the controversy over the Pleiades parallax as
measured by Hipparcos (van Leeuwen \& Hansen Ruiz 1998, Pinsonneault et al. 1998, Soderblom et al
1998) it seems wisest at present to remain with a
Hyades-based zeropoint.    Hipparcos has determined a very
accurate Hyades modulus of $3.33 \pm 0.01$ mag (Perryman et al. 1997) which so-happens to
agree exactly with the mean of all ground-based Hyades distance measurements made in
the past 20 years.  The often quoted  Feast \& Walker (1987) Cepheid ZAMS calibration was based
on a  Hyades modulus of 3.27, so their distance scale needs to be moved 0.06 mag more
distant, thus corresponding to an LMC modulus of 18.53.  A more 
recent discussion of the galactic ZAMS method by Laney \& Stobie (1994), where PL 
relations are derived in J, H and K as well as V, would move the distance scale only 
0.02 mag longer than this, if in both cases the Cepheids in associations 
are weighted half those in clusters and the same zeropoint is used, ie a LMC modulus
of 18.55 mag, and 18.96 mag for the SMC. 

The PL relation can also be calibrated directly using Hipparcos parallaxes of field
Cepheids.  Unfortunately with the exception of a single star (Polaris) these parallaxes
all have large relative errors.  Feast and Catchpole (1997) combine parallaxes for 26 Cepheids,
and determine a best fit $V$-band PL relation with zeropoint error $\pm 0.10$ mag.   With
LMC Cepheid photometry from Caldwell \& Laney (1991), reddening of $E(B-V) = 0.074$
and a metallicity correction of +0.042 mag, they derive an LMC distance modulus of 
$18.70 \pm 0.10$ mag.  As a caveat, the mean parallax of these 26 stars is only 2.1 ms arc.
Although the systematic error in the Hipparcos parallaxes is thought to be $\pm 0.1$ ms
arc (Brown et al. 1997), only a five percent effect for the Cepheids, treatment of systematic
effects at this level requires extraordinary care, and some caution in evaluation of
the results.

 Madore \& Freedman
(1998) compare the multi-wavelength LMC PL data (Madore \& Freedman 1991) with 
$BVIJHK$ photometry for the Hipparcos Cepheids, unfortunately only seven stars have 
mean magnitudes available in all six bands, thus the results are not very robust given the
large parallax errors for the galactic calibrators.  Discounting  their BV solution, which has 
very large error, and taking the mean of the remaining four solutions for various passband 
combinations, their LMC modulus is $18.54 \pm 0.12$.  Feast (1998) cautions that 
biases may be introduced by grouping the data in this way.

Fitting PL relations to MC data relies on having a large body of well-calibrated
light curves for the MC Cepheids.  The microlensing projects have produced
high quality light-curves for very large numbers of Cepheids, the potential of
which goes far beyond the use of Cepheids as distance indicators.  However
data in other important bandpasses (eg I, Tanvir 1996) and in the infrared is
less extensive than desirable.  In particular, Cepheids in the rich MC clusters
await definitive observations, although programs are now beginning to
address these needs (Ripepi 1998; W. Gieren, private communication).

\subsubsection {The Baade-Wesselink method}

The Baade-Wesselink (BW) method (Baade 1926, Wesselink 1946) and its
variant, the Barnes-Evans method (Barnes \& Evans 1976), can in principle provide
accurate distances to the MC Cepheids.  A very thorough description of the method
is given by Gautschy (1987), see also Balona (1977), and Feast \& Walker (1987), the
difficulties are succinctly summarized by Paczy\'{n}ski (1996).

In recent years the method, which
 requires preferably simultaneous photometry and radial velocity
measurements for the Cepheids, has moved to the infrared (Welch 1994, Laney \& 
Stobie 1995a,b, di Benedetto 1997, Gieren, Fouqu\'{e} \& Gomez 1998), resulting in a 
reduction in both systematic and random errors (compare Figs 6 - 11 of Fouqu\'{e} 
\& Gieren 1997).  Laney \& Stobie (1995a) summarize the infrared advantages: the light
variations at K are dominated by the change in surface area as opposed to 
temperature changes in the optical, infrared radius determinations are insensitive
to whether the phases corresponding to the ascending branch are excluded, and the
J-K or V-K color indices are insensitive to variations in microturbulence or
surface gravity throughout the pulsation cycle.

Calibration of the surface-brightness relations can be provided 
by a combination of model atmospheres and a  color-temperature calibration, but
a direct calibration is preferable.  It is now possible to use the many recent
interferometric angular diameter measurements of nearby giants and supergiants, 
a substantial fraction of these measurements are being made in the infrared thus
minimizing limb-darkening corrections.  Fouqu\'{e} \& Gieren (1997) have recently 
investigated in detail the applicability of the surface brightness - color 
relation for giants and supergiants to the Cepheids, and find excellent agreement 
with the slopes of the relations for all three types of star. Forcing the slope
to be that for the Cepheids, they determine very precise zero-points for the 
near-infrared surface brightness - color relationships.  Within the errors,
these zeropoints agree with those found via the Cepheid effective temperature 
scale (Pel 1978) and from the lunar occultation diameter of $\zeta$ Gem (Ridgway 
et al. 1982). 

Gieren  et al. (1998) note that their period-radius relation is identical to that found by 
Laney \& Stobie (1995b) despite the use of very different methods.  di Benedetto (1997)
adopts a steeper period-radius relation (slope 0.73 cf 0.68), more recent evidence (Bono, 
Caputo, \& Marconi 1998) favors the shallower value.
  
Model atmosphere analyses (di Benedetto 1997, Bell \& Gustafsson 1989) show that the
zero-point of the surface brightness - $(V-K)$ color relation 
is independent of metallicity to a level
much less than 0.01 mag over a range of at least 0.5 dex in $[Fe/H]$.  Perhaps the
major systematic uncertainty lies in the the {\em p} factor applied to the integrated
radial velocity curve in order to derive displacement, which is a correction for both
geometric projection and limb darkening effects.  

The results from all three recent investigations are in reasonable agreement.  Gieren et al
(1988) obtain an LMC modulus of 18.46 (they prefer to apply no metallicity correction, but
a correction of +0.06 is their suggested value), Laney \& Stobie (1995b)
find $18.58 \pm 0.04$ mag for the LMC and $19.00 \pm 0.04$ mag for the SMC, and di Benedetto (1997)
derives $18.64 \pm 0.02$ for the LMC and $19.06 \pm 0.03$ for the SMC, 
where for the latter two studies 
we have here increased the author's moduli by 0.06 so that the galactic cluster zeropoint 
corresponds to a Hyades modulus of 3.33 mag.  The zeropoint error is not included in those above.

Rather than use infrared BW observations to calibrate a PL relation, 
measuring BW radii for the MC Cepheids directly
would seem to be an attractive method for determining their distances in a very
straight-forward manner, since such a procedure is essentially metallicity and 
reddening independent (Gieren et al. 1998). Such programs are underway, for both for Cepheids in 
MC clusters (W. Gieren, private communication) and in M31 and M33 (Stanek et al. 1998).
LMC Clusters such as NGC 1866 and NGC 2031 each contain
many Cepheids and with the advent of large-format IR imagers at least the imaging observations 
can be made at high efficiency.

\subsection{RR Lyraes}

Udalski (1998) discusses distances to the MC and the galactic center based on observations
made in the course of the Optical Gravitational Lensing experiment (OGLE).  For 110
LMC and 128 SMC RR Lyraes he finds  mean $<I_{0}> = 18.41$ and 18.93 respectively, with
errors for each estimated as 0.02 mag (statistical) and 0.05 mag (systematic).  
Absorption corrections are $A_{I} = 0.33$ and 0.39 for the LMC fields, and
 $A_{I} = 0.16$ in 
the SMC.  Adopting a mean $(V-I)_{0} = 0.45$ and 0.48, then $<V_{0}> = 18.81$
and $<V_{0}> = 19.41$ for LMC and SMC respectively.  Similarly,  from the MACHO project database
(Alcock et al. 1998) mean properties are found for a sample of 3454 RRab variables,
the mean magnitude is $<V_{0}> = 19.00$.   The difference $\Delta V_{0} = 0.19$
mag between the two surveys could in principle almost all be due to the higher mean 
reddening adopted by Udalski (1998). 

These results can be compared with the mean
magnitudes of MC cluster RR Lyraes.  For 182 RR Lyrae in seven LMC clusters, 
the mean $<V_{0}> =
18.94 \pm 0.03$ (Walker 1992), while for four RR Lyraes in the SMC cluster NGC 121,
Walker \& Mack (1988) find $<V_{0}> = 19.46 \pm 0.07$.  The SMC comparison should
not be over-interpreted given the uncertain location of NGC 121 with respect to the
SMC center, although NGC 121 {\it does} appear to be located in a region of the SMC with
relatively small depth in the line of sight (Gardiner \& Hatzidimitriou 1992).
 The mean metallicity of the cluster RR Lyraes is near $[Fe/H] = -1.9$ (Walker 1992)
while that for the field stars is rather uncertain, for instance from
a period-amplitude analysis Alcock et al. (1998) find a mean metallicity of $[Fe/H] = -1.2$,
rather more metal rich than the canonical $[Fe/H] \sim -1.6$ often assumed.  With
a slope of $\sim 0.2$ for the RR Lyrae magnitude-metallicity relation 
(Fernley et al. 1998a) the cluster stars
are expected to be $0.08 - 0.16$ mag brighter than the field stars on this basis.
Given that the MC clusters with RR Lyraes have generous numbers of variables compared
to the non-variable horizontal branch population, the majority of the cluster variables
are expected from evolutionary lifetime arguments to be close to the ZAHB, and thus in
this respect similar to the field population.  Therefore the Walker (1992) and Alcock et 
al (1998) results seem consistent but the Udalski (1998) stars appear to be too 
bright by comparison. 

To proceed further requires a calibration of the absolute magnitudes of RR Lyraes.
This is controversial.  Statistical parallax (Layden et al. 1996, Popowski \& Gould 1997,
Fernley et al. 1998b) and 
Baade-Wesselink  (Carney et al. 1992, Clementini et al. 1995) analyses of galactic field RR 
Lyraes find them fainter by typically $0.2-0.3$ mag than calibrations based on subdwarf parallaxes, 
(Gratton et al. 1997, Reid 1997, Pont et al. 1998), evolution theory 
(Caloi, D'Antona, \& Mazzitelli 1997) and pulsation theory applied to the 
double-mode (RRd) variables (Alcock et al. 1997a, Kov\'{a}cs \& Walker 1998).  All the 
latter suggest an LMC modulus of $\sim 18.5$ mag.  Catelan (1998) showed that
galactic cluster and field RR Lyraes have the same distribution in the period-temperature
diagram, and he argues that any a consequence of this result is that any difference in
luminosity between the two groups of stars is very unlikely.  The LMC results above further
support this result.  Only a single star, RR Lyrae itself, has a Hipparcos parallax of
any significance, with a consequent $\pm 0.3$ mag error in its absolute magnitude (Fernley
et al. 1997).

\subsection{Mira Variables}

The use of Mira variables as distant indicators is discussed in detail elsewhere in this
volume.  Occuring in the general field, in metal rich globular clusters, and in the galactic 
center, as well as being easily bright enough to be accurately measured in 
local group galaxies, they are an important complement to more traditional 
distance indicators such as the Cepheids and RR Lyraes.  
Infrared PL relations with small scatter have been found for Mira variables in the MC 
(Feast et al. 1989, Groenewegen \& Whitelock 1996).  Wood (1995) found no strong evidence
for a metallicity dependence, by comparing results for LMC and SMC Miras, but earlier
Wood (1990) had suggested that $M_{K}$ should be less sensitive to metallicity effects than
$M_{bol}$.  The zeropoint can be calibrated from Miras in metal rich globular clusters,
these with distances by other means (eg RR Lyraes, main sequence (MS) fitting to local subdwarfs),
from an assumed distance to the galactic center or, most directly, from Hipparcos parallaxes
to a few nearby Miras.  As with the Cepheids, there are few such stars in the catalog.
Restricting the
sample to the 11 oxygen-rich Miras with Hipparcos parallaxes, and defining the PL relation
slope from the LMC Miras, a mean of the $M_{K}$ and $M_{bol}$ relations gives an LMC modulus of
$18.54 \pm 0.18$.  If $M_{K}$ alone is to be preferred then the distance increases to $18.60
\pm 0.18$.

\subsection{Detached Eclipsing Binaries}

Paczy\'{n}ski (1996) has reviewed the use of detached eclipsing binary systems as
distance indicators.  He advocates the use of double-lined systems as distance
indicators, applicable to galaxies throughout the Local Group.   For the MC,
many potential candidates have been identified from the microlensing surveys (eg
Alcock et al. 1997b).
Although intensive observing is then needed to obtain accurate photometry and radial 
velocities, the method has the advantage of being near-direct.  If 
the selected binary is indeed well-detached and uncomplicated, then only a surface
brightness  - color relation is needed in order to calculate the distance, in addition to
directly measured quantities.  This relation can be calibrated from interferometrically
measured stellar angular diameters, and as for the Cepheids, it is probably best to use a
color index such as $V-K$.   Guinan et al. (1997) provide a preliminary report on 
results for the LMC eclipsing binary HV 2774, combining ground-based photometry and 
spectroscopy with HST spectroscopy.  They find a distance modulus for the LMC of 
$18.54 \pm 0.08$ mag.

\subsection{Other Methods}

Several other methods can provide distances to the MC.  The tip of the red giant 
branch (TRGB), (Lee, Freedman \& Madore 1993, Madore, Freedman \& Sakai 1996) can be
clearly defined given sufficient numbers of stars, and appears to be an excellent
distant indicator for low-metallicity populations, particularly in an $I, V-I$ CMD
where the externally defined dispersion is less than $\pm0.1$ mag.
Since it relies on the galactic globular cluster for calibration, the TRGB method
is subject to the same distance scale uncertainties that plague the RR Lyrae
distance scale.  

The Planetary Nebulae (PN) luminosity function (Jacoby et al. 1992, Jacoby 1997) 
is calibrated by assuming
a distance to M31.  The Jacoby, Walker \& Ciardullo (1990) distance moduli for the MC, adjusting 
to the Freedman \& Madore (1990) M31 modulus, are $18.50 \pm 0.18$ for the LMC, based on 42 PN,
and $19.15 \pm 0.29$ for the SMC, based on 8 PN.  Since the M31 distance is based on a 
galactic Cepheid calibration, the PN do not provide independent zeropoints for the MC.

Bond (1996) describes the use of post asymptotic giant-branch (PAGB) stars as distance 
indicators.  The galactic calibration on such stars in galactic globular clusters, and so
again is tied to the RR Lyrae distance scale, together with its present calibration uncertainties.
Given the rather few PAGB galactic calibrators, the main use of such stars in the MC 
is likely to be in strengthening the calibration, once a definitive distance to the MC can be
found by other means.

Various luminosity calibrations of novae light curves have been presented by Della Valle \&
Livio (1995) and Livio (1997), based mostly on novae in M31, but also
include 15 LMC novae.  Livio (1997) lists the several advantages of novae as distance
indicators which includes their brightness, no metallicity dependence, good theoretical 
understanding, and rather small intrinsic scatter.   Their discovery and consequent study
is observationally intensive, and like the PN and PAGB stars, the use of the MC novae 
is more important in strengthening the galactic calibration by assuming MC distances from
some other source.
 
Many color magnitude diagrams (CMDs) of MC clusters appear in the literature, and in the 15 years since
the introduction of CCDs the resulting photometry has been accurate enough to allow distances
to be derived either by comparison either with similar galactic clusters, assuming a distance for the
latter, or with theoretical isochrones.  The isochrone comparison has traditionally been used to determine
all of age, distance, metallicity and reddening, a sufficient number of variables that it
is near-impossible to test any of the assumptions (mixing-length, overshoot, etc.) used in
building the isochrones.  Far better is to measure metallicity and reddening separately so
that the relevance of the chosen set of isochrones can be definitively tested.  The many MC
intermediate age clusters are obvious targets, the younger of these may 
be able to be compared directly to the Hyades, with suitable differential metallicity corrections.   
For most of the clusters, spectroscopic metallicities are not known.
Available CMDs (see list in Westerlund 1997) have favored LMC distance moduli in the range
$18.4 - 18.6$, early smaller moduli are mostly a consequence of incorrect isochrone color-temperature
calibrations.   The younger populous MC clusters, such as NGC 1866, are also prime targets, with
the advantage that several contain significant numbers of Cepheid variables thus allowing a direct
distance comparison.  If CMD-based distances accurate to $<0.1$ mag are to be produced then the  
absolute photometric accuracy demanded of the observations is very high due to the steepness of the main
sequence in the CMD. Dereddened colors of the MS should have error no larger than $\pm0.02$ mag, 
and preferably nearer to $\pm0.01$ mag.  It is doubtful whether any of the published CMD's have 
reached this level of accuracy, although it is certainly not an impossible task given sufficient
attention to the calibration issues.

\section{CONCLUSIONS, AND THE FUTURE}   

The Hipparcos mission has provided parallaxes for the
traditional distance indicators such as Cepheids and Miras, enabled 
alternative calibrations for the open cluster 
route to Cepheid luminosities and via subdwarfs to Miras and RR
Lyraes in globular clusters.  It has also allowed the invention
of new indicators, such as the red clump stars.  All these distance
indicators are present in the MC, where the luminosity scales 
can be compared directly.  Although the mission was undeniably
a great success, it is unfortunate that the accuracy limits to the 
Hipparcos parallaxes are such that the Cepheid, RR Lyrae and Mira 
distance scales cannot be fixed to the few percent accuracy that
is scientifically so desirable.  

Microlensing surveys have provided high quality optical photometry for
many thousands of variables in the MC, the galactic bulge and a few
other selected targets.  The great value of this photometry for the
pulsating variables is that it provides statistically significant samples
of stars with differing masses, temperatures, metallicities and pulsation 
modes in order to compare with evolution and pulsation theory.  Our
understanding of these stars will be greatly improved as a result,
which should enhance their value as distance indicators.

Distance measurements that contain the minimum number of steps and
assumptions will be those subject to the least number of systematic
errors.  
The SN 1987A distance to the LMC is one that is independent
of all other distance indicators, and as discussed above, more definitive
quantification of possible systematic effects should be possible from
observations made in the next few years.   Eclipsing binaries have been
discovered in profusion in the MC by the microlensing surveys, and 
detached double-lined systems can provide a near-direct distance estimate.
The direct infrared Baade-Wesselink calibration of MC Cepheids is now 
also underway, with increased confidence that the systematic effects that
have plagued the visual wavelengths version of the method are indeed
under control.

The new large telescopes in the south (ESO VLT, Gemini, Magellan) together
with new instrumentation, particularly that working in the near-infrared, will
play a large part in resolving MC distance concerns, with accurate observations
of many of the distance indicators mentioned above,  and will provide
detailed spectroscopic analyses of MC populations in general.  The move to the infrared
has reduced dependence on reddening, but definitive calibrations of
metallicity effects are not yet available.

Our present evaluation is that the center of the SMC has a modulus $0.42
\pm 0.05$ greater than that of the LMC, a result depending mostly on the Cepheids.
It assumes for both LMC and SMC that the space distributions of the two sets of 
stars are not, in the mean, offset from the centers of the galaxies.
A ``best'' distance for the LMC can be obtained either by taking a simple
mean of all distance indicators, or else by suitable weighting of the indicators.
Since systematic errors are likely to dominate in almost all cases, the former
method is rather unsatisfactory due to the possible dominance of outliers, while
the latter relies on ad-hoc weighting that is very difficult to quantify
scientifically.  The evaluation here is that calibrations based on
galactic field RR Lyraes, RGB clump stars, and comparison of MC cluster CMD's to
isochrones or galactic clusters, should be weighted zero at present.
That leaves us with SN 1987A, Cepheids (via clusters, Baade-Wesselink, Hipparcos),
RR Lyraes (via Hipparcos subdwarf calibration of Globular cluster distances), 
Miras, and an eclipsing binary. 
For these a mean modulus is $18.55 \pm 0.10$ mag, where the error estimate is
approximate and one which conservative readers versed in the history of the
subject may well prefer to double.   A different weighting scheme can produce
very different results, for instance (Fernley et al. 1997), emphasising 
a calibration based on galactic RR Lyraes and discounting the distance indicators
prefered here produces an LMC modulus near 18.3 mag.  Until discrepancies such
as this are sorted out we cannot be {\it entirely} sure that we have made the
correct choice.

It is clear from the discussions in this chapter that we can only
conclude, as did Westerlund (1997), that the distances to the MC are not
yet sufficiently well-known, despite the success of the Hipparcos mission and
the invaluable microlensing photometry.  Further analyses of their results,
and the results of on-going programs should over the next few years provide
more definitive distances.  In the longer term, beyond 2005, the Space Interferometry
Mission (SIM, see http://sim.jpl.nasa.gov/sim/) is expected to provide parallaxes 
accurate to 4 $\mu$arcsec for
10000 stars, and which corresponds to a $5\sigma$ measure for a star in the LMC,
while the later GAIA mission (Lindegren \& Perryman 1996) will provide 50 million
parallaxes with an accuracy of better than 10 $\mu$arcsec.  These missions will
certainly resolve all questions relating to distances within our galaxy, and 
reduce the uncertainty in the MC distances to below one percent.

{}
\end{document}